\documentclass[structabstract,a4paper]{aa}
\usepackage{natbib}
\usepackage{graphicx,txfonts,latexsym}
\usepackage{subfigure}
\usepackage{hyphenat}
\usepackage[latin1]{inputenc}
\usepackage{booktabs}

\bibpunct{(}{)}{;}{a}{}{,}

\begin{document}

\title{A Neural Network Gravitational Arc Finder based on the Mediatrix filamentation method}


\author{C.~R.~Bom\inst{1,2}
\and M. ~Makler\inst{1}
\and M.~P.~Albuquerque\inst{1}
\and C.~H.~Brandt\inst{3,4,5}}

\institute{Centro Brasileiro de Pesquisas F\'isicas, Rua Dr. Xavier Sigaud 150, CEP 22290-180, Rio de Janeiro, RJ, Brazil
\and Centro Federal de Educa\c{c}\~ao Tecnol\'ogica Celso Suckow da Fonseca, Rodovia M\'ario Covas,
lote J2, quadra J, CEP 23810-000,  Itagua\'i, RJ, Brazil 
\and International Center for Relativistic Astrophysics Network (ICRANet), University of Rome ``La Sapienza'', 
P.le Aldo Moro, 5, 00185 Roma, Italy 
\and 
ASI,  Via del Politecnico snc, 00133 Roma, Italy, 
\and
CAPES Foundation, Ministry of Education of Brazil, CEP 70040-020, Brasilia, DF, Brazil}

\abstract
{Automated arc detection methods are needed to scan 
the ongoing and next-generation wide-field imaging surveys, which are expected to contain thousands of strong lensing systems. Arc finders are also required for a quantitative comparison between predictions and observations of arc abundance. 
Several algorithms have been proposed to this end, but machine learning methods have remained as a relatively unexplored step in the arc finding process.}
{In this work we introduce a new arc finder based on pattern recognition, which  uses a set of morphological measurements that are derived from the Mediatrix filamentation method as entries to an artificial neural network (ANN).  
We show a full example of the application of the arc finder, 
first training and validating the ANN on simulated arcs 
and then applying the code on four Hubble Space Telescope (HST) images of strong lensing systems.}
{The simulated arcs use simple prescriptions for the lens and the source, while mimicking HST observational conditions. We also consider a sample of objects from HST images with no arcs in the training of the ANN classification.
We use the training and validation process to determine a suitable set of ANN configurations, including the combination of inputs from the Mediatrix method, so as to maximize the completeness while keeping the false positives low.} 
{In the simulations the method was able to achieve a completeness of about $90\%$ with respect to the arcs that are input into the ANN after a preselection. However, this completeness drops to $\sim 70\%$ on the HST images. The false detections are on the order of $3\%$ of the objects detected in these images.}
{The combination of Mediatrix measurements with an 
ANN is a promising tool for the pattern-recognition phase of arc finding. More realistic simulations and a larger set of real systems are needed for a better training and assessment of the efficiency of the method.}

\keywords{Gravitational lensing: strong, Techniques: image processing, Methods: numerical}

\titlerunning{Mediatrix Arcfinder} 
\authorrunning{C.~R.~Bom et al.}
\maketitle

\section{Introduction \label{introd}}

Strong lensing  
provides a useful tool to uncover the mass distribution in
galaxies \citep[e.g.,][]{0004-637X-575-1-87,2002MNRAS.337L...6T,2006ApJ...649..599K} 
and galaxy clusters \citep[e.g.,][]{1989ApJ...337..621K,1998MNRAS.294..734A,2007MNRAS.376..180N,2010AdAst2010E...9Z,2010ApJ...715L.160C,2010ApJ...723.1678C}.
Gravitational arcs have also been used to constrain the background cosmological model
\citep[e.g.,][]{1998A&A...330....1B,1999A&A...341..653C,2002A&A...387..788G,2002MNRAS.337L...6T,2001PThPh.106..917Y,2004MPLA...19.1083M, 2005MNRAS.362.1301M,2010Sci...329..924J,Magana2015,0004-637X-806-2-185,2016A&A...587A..80C}.
More recently, arcs and Einstein rings, in combination with kinematic information for the lenses, have been employed for testing modified gravity \citep[e.g.,][]{Schwab2010,Enander2013,2016arXiv160203385P}.
Furthermore, 
strong lenses are also being exploited as cosmic telescopes, enabling spectroscopic and spatially resolved studies of high-redshift sources, such as
dwarf galaxies at distant redshifts \citep{Marshal2007},  star-forming galaxies \citep{Stark2008}, quasar accretion disks \citep{2008ApJ...673...34P}, and faint Lyman-alpha blobs \citep{2015arXiv151205655C}.


The many applications of gravitational arcs in astrophysics and cosmology have spurred the search for these objects in both space-based and ground-based observations. 
This includes searches in Hubble Space Telescope (HST) mosaics,  such as the {\it Hubble Deep Field}  \citep[HDF;][]{1996ApJ...467L..73H}, {\it HST Medium Deep Survey} \citep{1999AJ....117.2010R}, {\it Great Observatories Origins Deep Survey} \citep[GOODS;][]{2004ApJ...600L.155F}, 
 {\it Extended Groth Strip} \citep[EGS;][]{Marshall2009robot},  {\it HST Cosmic Evolution survey} \citep[COSMOS;][]{2008ApJS..176...19F,2008MNRAS.389.1311J}
and in targeted observations of galaxies 
\citep{2006ApJ...638..703B,2012ApJ...744...41B}
and clusters \citep{2005Smith,2005ApJ...627...32S,2010MNRAS.406.1318H,2016ApJ...817...85X}. 
Investigations from the ground include 
follow-ups of clusters 
\citep{1999A&AS..136..117L, 2003ApJ...584..691Z,2008AJ....135..664H,2010A&A...513A...8K,2013MNRAS.432...73F} 
and galaxies \citep{2006MNRAS.369.1521W}, 
and searches in 
wide-field surveys, such as the {\it Red-Sequence Cluster Survey} \citep[RCS;][]{2003ApJ...593...48G,2012ApJ...744..156B},
{\it Sloan Digital Sky Survey} \citep[SDSS;][]{2007Estrada,2009MNRAS.392..104B, 2010ApJ...724L.137K, 2011RAA....11.1185W,2012ApJ...744..156B},  {\it Deep Lens Survey} \citep[DLS;][]{Kubo2008}, 
 {\it  Canada-France-Hawaii Telescope (CFHT) Legacy Survey} 
\citep[CFHTLS;][]{2007A&A...461..813C, 2012More,Maturi2014,RINGFINDER,SPACEWARPSII,ParaficzCFHTLS}, 
 {\it CFHT Stripe 82 Survey} (CS82; Caminha, More et al., in prep.), and  Dark Energy Survey\footnote{\texttt{http://www.darkenergysurvey.org}} \citep[DES;][]{DES01} Science Verification data \citep{DESSL1Nord}.

As of now, the largest homogeneous samples of gravitational arcs have on the order of a hundred systems. These numbers will increase by one order of magnitude with the close completion of the Kilo Degree Survey\footnote{\texttt{http://kids.strw.leidenuniv.nl/}} \citep[KiDS;][]{KiDSr2} and DES \citep{DESnonDE}, which will cover, respectively, 1000 and 5000 square degrees with sub-arcsecond seeing. 
Comparable numbers are expected from the ongoing Hyper Suprime-Cam\footnote{\texttt{http://www.naoj.org/Projects/HSC/surveyplan.html}} (HSC) and the forthcoming Javalambre Physics of the Accelerating Universe Astrophysical Survey \citep[J-PAS;][]{JPASredBook} projects.
These numbers are expected to increase even further in the near future, with the operation of the Large Synoptic Survey Telescope \citep[LSST;][]{LSST20} and 
Euclid\footnote{\texttt{http://www.euclid-ec.org/}} 
\citep{EuclidSciBook},
which are both expected to detect $\mathcal{O}\left(10^5\right)$ systems with arcs \citep{Collet2015}.

The vast majority of the current samples of arc systems involve a visual search and classification. This is true for the targeted surveys and also for the wide-field imaging surveys, where either the full footprint or cutouts 
around potential lenses (e.g., luminous red-galaxies, galaxy clusters) are visually inspected. 
This manual procedure is still possible for current surveys with good image quality,
which cover at most few hundred square degrees. However, it will become prohibitive for DES and KiDS, and especially for LSST and Euclid. Therefore, the development of automated arc finding methods is absolutely needed for the scrutiny of these surveys in the quest for gravitational arcs.

Regardless of the size of the survey, automated arc detection is important for an objective and reproducible definition of arc samples, which often includes the determination of arc properties. 
This is of course critical for arc statistics \citep[see, e.g.,][]{2013SSRv..177...31M,2016ApJ...817...85X} and for any comparison of real and simulated data \citep[e.g.,][]{2005Horesh,2011MNRAS.418...54H}
and among different data sets  \citep[e.g.,][]{2010MNRAS.406.1318H}.


Motivated by these needs, several automated methods to find gravitational arcs have been proposed in recent years.
Most of these methods focus on pattern recognition, i.e., on identifying shapes that look like gravitation arcs, in particular, thin and elongated structures \citep[e.g.,][]{2004Lenzen,2005Horesh, 2006Alard,2007Seidel, 2012More}; 
in some cases, these methods require a degree of curvature 
\citep[e.g.,][]{2007Estrada,Kubo2008}. 
\citet{Maturi2014} combine this approach with a multicolor selection of the sources.
\citet{Marshall2009robot} use lens inverse modeling to find strong lenses, i.e.,  assuming that a given object in an image is a consequence of lensing and determining whether the lensing solution is favored by the data. 
More recently, new arc finders have been proposed that subtract the lens candidate (usually early-type galaxies) light distribution,  either using two bands, as in \citet{RINGFINDER}, or by modeling the lens in a single band, as in \citet{JosephPCA2014} and \citet{Brault2015}.
The residuals are then investigated, using  their shapes \citep{JosephPCA2014}, by color selection \citep{RINGFINDER}, or with inverse modeling \citep{Brault2015}.

The inverse modeling approach is particularly interesting as it uses the physics of lensing to find candidates, however its applicability has been restricted to galaxy-scale lenses; this is because the automated modeling is much more tractable in this case owing both to the simplicity of the lens model and the identification of the images. 
The lens subtraction approach is a necessity for galaxy-scale lenses, especially when observed from the ground, as the arcs can be embedded in the galaxy's light. 
On the other hand it is less critical for arcs on cluster scales, which span larger angular sizes than the galaxies and the PSF.

Our main interest here is to look for arcs in galaxy groups and clusters that can be found even with a single band survey (as in CS82). Therefore, in this paper we focus only on the development of a pattern-recognition based arc finder.
Most arc finders in this category use sets of measurements of the objects, such as ellipticity, length, $L$, width $W$, and {\it arcness,} to determine whether they are arc candidates or not. They usually employ hard (i.e.,  fixed and mutually independent) cuts, whose values may be arbitrarily assigned or tuned using data or simulations. However, given the diversity of arc properties (shapes, sizes, $S/N$ ratios, etc.) and their physical origin, different 
cuts could perform better in different regions of the multidimensional space of arc parameters. 
For example, arcs may be very elongated and not necessarily curved for galaxy  cluster lenses, while arcs are not as drastically elongated for galaxy lenses but exhibit a clear curvature.
Therefore, a flexible criterium based on a combination of parameters may be more efficient than applying hard cuts.
This is a typical situation in which machine learning methods can be extremely helpful. A suitably trained algorithm can then classify the objects into arcs or not, given a set of input values for the object features. Such training can be carried out either on real data (on objects previously known to be arcs) or using simulations, by feeding the algorithm with a large  set of arc and nonarc samples. This process is 
characteristic of supervised learning methods, of which the most well known is the back-propagation artificial neural network \citep[ANN;][]{williams1986learning,rumelhart1988learning}.

The choice of the set of input parameters is as important as the choice of the classification method and its configurations. In this work we adopt measurements derived from the Mediatrix filamentation method \citep{2012Bom,BomMediatrix},  a novel iterative technique that decomposes elongated objects into segments along their intensity ridge line. This method provides several morphological parameters that are well suited to characterize arcs, including the length along the ridge line, $L$, the width $W$, and, most notably, estimates of the object center of curvature and its significance.

Therefore, the purpose of this work is to construct an ANN gravitational arc finder based on the Mediatrix filamentation method, or {\it ANN Mediatrix Arcfinder} (AMA) for short. We use a sample of simulated gravitational arcs and a sample of nonarcs from HST images to train and validate the ANN. This sample is used to 
pin down a few configurations among the many possible choices involved in the ANN detection process: the types of images used for the training, the selection of inputs given to the ANN, the number of neurons, and the final threshold for classification. 
As an illustration of the application of the method to real data, we consider four galaxy cluster images from HST and run the AMA on them, comparing the results with the training and validation.

The paper is organized as follows. Next section provides an overview of the AMA algorithm: the processes to detect and select the objects, the measurements carried out on them, in particular the  Mediatrix filamentation, and the ANN used in the arc identification. 
Section \ref{train_and_validation} describes the training and validation process, the samples used, and the tuning of the ANN for arc detection. Section \ref{aplication}  shows an example of application of the method on real data.
Final discussions and concluding remarks are presented in section
\ref{discusao}.

\section{\label{apresentacao} The Mediatrix arcfinder algorithm}

Most gravitational arc finding methods based on pattern recognition techniques can be divided into four steps: object
segmentation, preselection, measurement, and final classification. 
In the segmentation phase, sets of pixels above the background are grouped into objects (as discussed in Sec. \ref{identificacao}).
In the preselection phase, we define a sample of objects to be analyzed, performing cuts to eliminate those that can be readily discarded as not being arcs (Sec. \ref{preselection}). 
The measurements are carried out through 
the Mediatrix filamentation method \citep[][see Sec. \ref{mediatrix}]{2012Bom}, which provides a number of parameters that characterize the objects.  
For the final classification
we use an ANN trained to identify arc candidates (see Sec. \ref{RNA}). 

\subsection{ \label{identificacao} Object segmentation} 

The first step is to identify the objects in the image, separating them from the background and defining which set of pixels belong to a given object. To this end, we use the \texttt{SExtractor} \citep{bertin96} software, which has several parameters controlling the object identification, deblending, and measurement process, including the minimum signal-to-noise ratio for a given pixel to be considered, minimum number of pixels, and deblending thresholds.
Tuning these parameters is a key step in arc identification, especially as arcs are low surface brightness objects, are often close to brighter sources, and can thus be easily missed and/or blended with other sources.
However, in this paper we do not perform a systematic optimization of these parameters. We rather use a set of manually tuned values that provided good results on a visual inspection, as our main focus is in the measurement and final classification phases. 
 \citet{2005Horesh}, \citet{2007Estrada}, and \citet{Kubo2008} also use \texttt{SExtractor} in the object  segmentation phase of their arc finding methods.

\texttt{SExtractor} provides an output catalog containing measurements on the objects identified and several image outputs with the same dimensions as the original one. Here we use two such output images, namely \texttt{OBJECTS}, containing the pixel values of all objects identified, and \texttt{SEGMENTATION}, in which all pixels belonging to the same object have the same value (corresponding to the object ID in the catalog). From these two images we produce a single array per object, which contains only the pixels belonging to that object and the respective pixel value. These arrays are called {\it postage stamps} and are kept in the memory for the next steps. From this point on the AMA algorithm will work separately on each object. 

\subsection{ \label{preselection} Preselection}

Among the measurements provided by \texttt{SExtractor} are the object semimajor axis  $A$ and 
semiminor axis $B$ derived from the weighted central second moments of the pixels \citep{bertin96}.
From them we define the
ellipticity $e$ as
\begin{equation}
e=1-\frac{B}{A}, 
\end{equation}
which is used to eliminate from the sample objects with ellipticities below some threshold $e_{\rm th}$.
For the images used in this paper, we set $e_{\rm th} = 0.4$.
We also add a cut on the maximum number of pixels to exclude objects that are too large and are definitely nonarcs.
To avoid spurious detections we remove objects that are close to the image borders.
We do not make cuts in the object signal-to-noise ratio or magnitude, as this could remove some of the faint arcs. 
We do not apply any star-galaxy separation either, as the cut on $e$ already removes the stars.

\subsection{\label{mediatrix} Measurements with the Mediatrix filamentation method}

In this section we review the Mediatrix filamentation method \citep{2012Bom,BomMediatrix} and introduce the outputs of this process that are used to produce the inputs for the ANN to identify the arcs.
The Mediatrix filamentation method is an iterative algorithm to assign $N$ filaments along the intensity ridgeline of a given object and derive morphological measurements such as length, width, curvature, and curvature center.

\begin{figure}[!ht]
 \begin{center}
 \sidecaption
 \resizebox{\hsize}{!}{\includegraphics{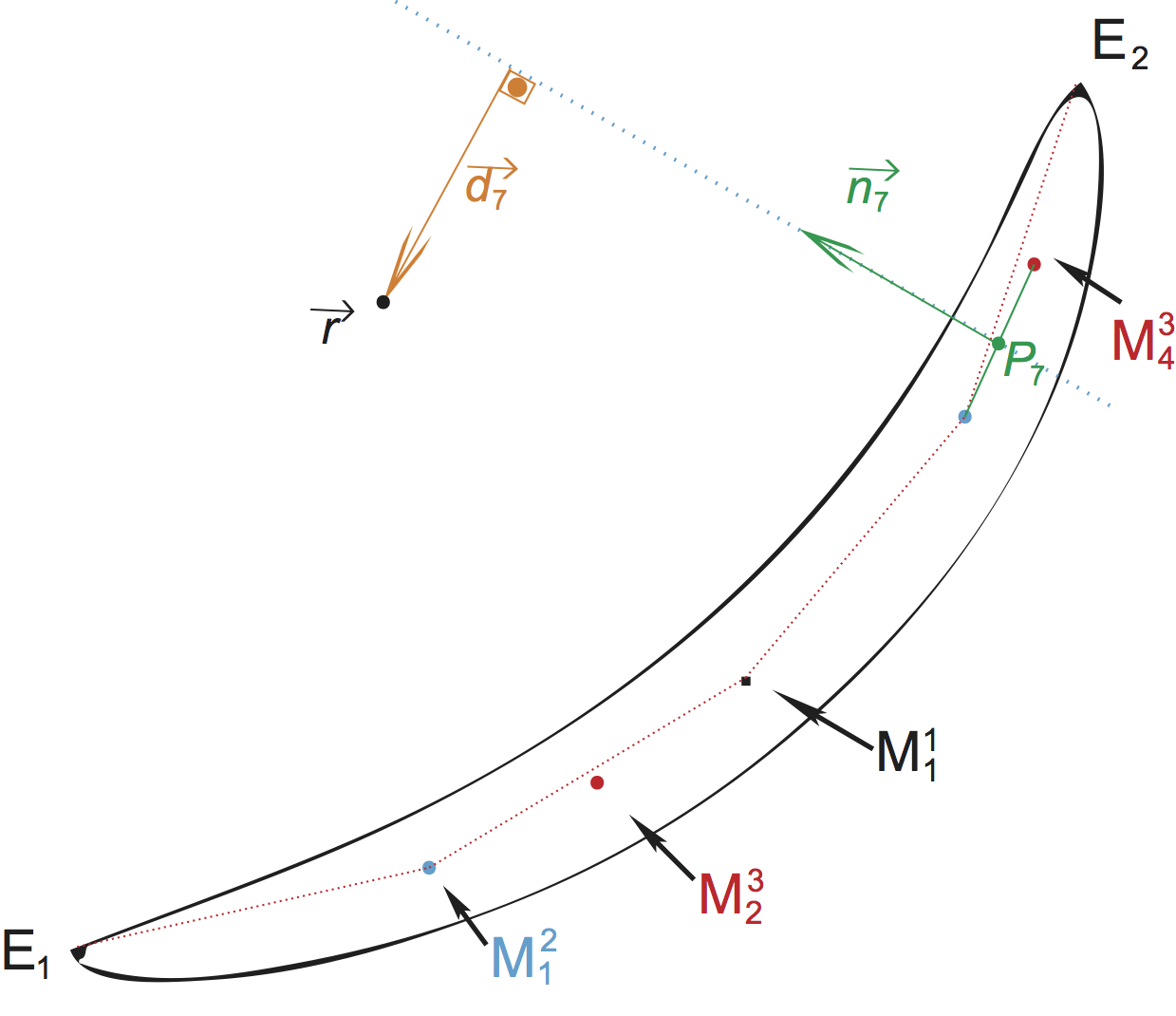}} 
 \caption{\label{d_definition}
Illustration of the Mediatrix filamentation points, corresponding to an iteration level of $n=3$.  Only one vector and a few points and segments are shown for clarity.  The object chosen is an ArcEllipse \citep{arcellipse} for illustration purposes. Also shown is the distance $\vec{d}_j$ from a point $\vec{r}$ on the plane to the line spanned the vector $\vec{n_i}$.  }
 \end{center}
 \end{figure}

\begin{figure}[!ht]
 \begin{center}
 \sidecaption
 \resizebox{\hsize}{!}{\includegraphics{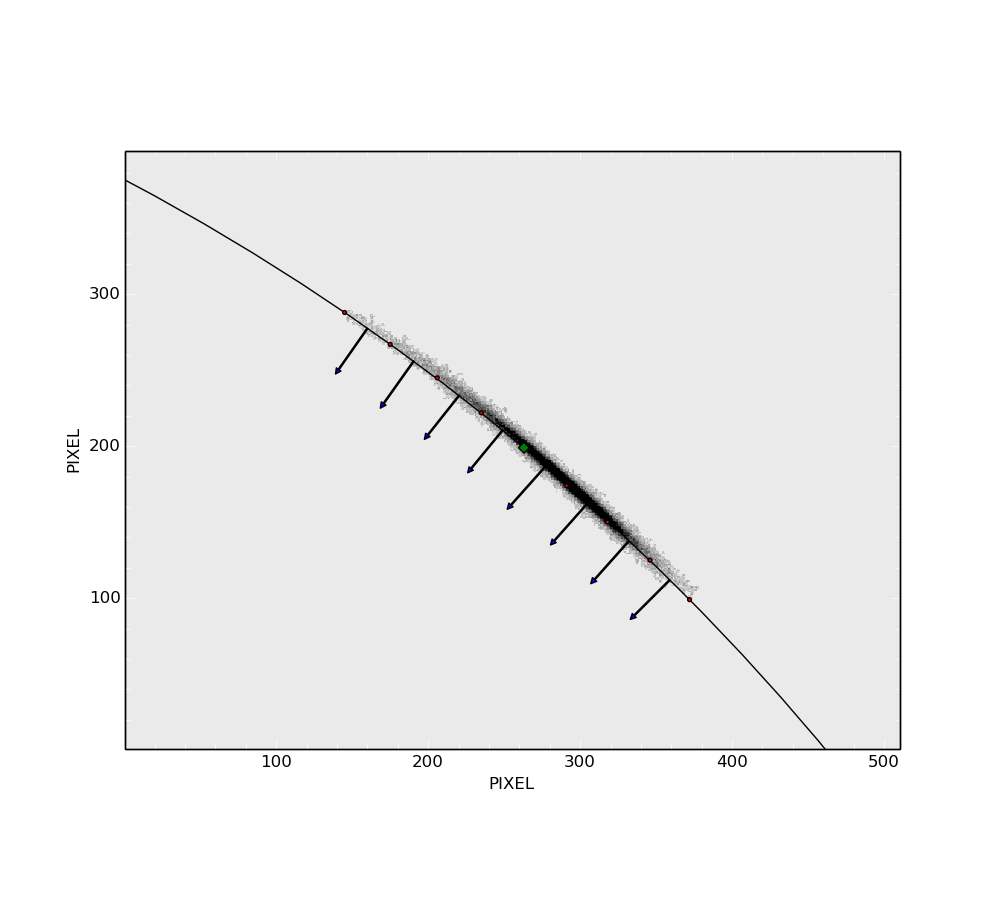}}
 \caption{\label{mediatrixexample}
Example of the Mediatrix method applied to a simulated arc for $n=3$. The dots show the Mediatrix points and the arrows the $\vec{n_i}$ vectors. The solid line is the circle that contains points $E_1$, $M^1_1$, and $E_2$.
}
 \end{center}
 \end{figure}

The method operates on a set of points with a given intensity, i.e., in the postage stamps of the objects after the preselection phase. The first step is to find the two most distant set of points $E_1$ and $E_2$. The first Mediatrix point $M^1_1$ is defined by the highest intensity pixel of the object along the perpendicular bisector to the segment $\overline{E_1E_2}$ (see Fig. \ref{d_definition}). In practice this is chosen as the highest intensity pixel from all pixels with positions $\vec{x}_i$ whose  distance $s_{ij}$ to the bisector $y_{j}$ satisfies the condition: $s_{ij}(\vec{x}_i,y_j) \leq \frac{\sqrt{2}}{2}dpix$, where $dpix$ is the pixel scale.
The same procedure is repeated between $E_1$ and $M^1_1$ and $E_2$ and $M^1_1$, defining a new pair of Mediatrix points, $M^2_1$ and $M^2_2$, where the superscript defines the iteration level and the subscript denotes each point. The whole process is repeated $n$ times. For each step $k$, the pair of neighboring points from all previous steps defines a new Mediatrix point, $M^{k}_{l}$, the $l$-$est$ point in the $k$-$est$ iteration level.
The output of the Mediatrix method is a set of
$N$ vectors $\vec{n_i}$, with $|\vec{n_i}|=l_i$, where $l_i$ is the distance between a given pair of neighboring Mediatrix points. The orientation is  
perpendicular to $l_i$ and its origin in the middle point of $l_i$. An example of the output of the Mediatrix filamentation on a simulated gravitational arc is shown in Fig. \ref{mediatrixexample}. 

For an iteration level $k$ we define the object length $L_k$ as
\begin{equation}
\label{length} L_k= \sum^{N}_{i=1} l_i.
\end{equation}  
and width $W_k$ as 
\begin{equation}
\label{width}
W_k=\frac{4A}{\pi L_k}, 
\end{equation}
where $A$ is the object area in pixels.
This relation between area, length and width is exact for arc-shaped objects known as ArcEllipses \citep{arcellipse}, as used to illustrate Fig. \ref{d_definition},
and is an excellent approximation for a specific solution for gravitational arcs (Pacheco et al. in preparation).

When the segments  $l_i$  are too small, there is no point in continuing the Mediatrix iteration. On the contrary, the decomposition starts to be dominated by noise (or by the finite pixel size). We expect that when $l_i \lesssim W_k$ the directional information is lost and no further division is useful. Therefore we impose the following condition as a criterium to stop the iteration:
\begin{equation}
\label{limite}
l_i \leq \alpha W_k, 
\end{equation}
where 
$\alpha$ is a parameter that
 is set to $\alpha=1$ in this work.

The decision to continue with the Mediatrix decomposition is taken independently for each segment following Eq. (\ref{limite}).  The iteration generally stops for regular objects at the same level for all segments, such that the total number of segments (and oriented vectors) is $N=2^n$, where $n$ is the last iteration level. However, if the shape is irregular, the iteration can be carried out to different levels for different regions of the object. 
Also, if the object is composed by noncontiguous sets of pixels, the code may not find any pixel along the perpendicular bisector of two given Mediatrix points. In this case, the iteration is stopped so that no further division happens between those two points.
After the last iteration, the sum in Eq. (\ref{length}) is carried out for all segments defining the final length $L$. 
The final width $W$ is defined using Eq. (\ref{width}).

We may define a center of curvature by determining the circle that passes through the points $E_1$, $M^1_1$, and $E_2$ (as shown in Fig. \ref{mediatrixexample}). The center of this circle, $\vec{r}_c$, is defined as a center of curvature and the curvature radius is the circle radius, $R_c$. Using these points we may also define a length as the circle arc length, $L_c$, between the two extreme points, $E_1$ and $E_2$. 

Another possibility is to define a curvature center based on all points from the Mediatrix filamentation: the point on the plane that minimizes the distances to the perpendicular bisectors of all Mediatrix segments $l_i$ (i.e., the distance to the lines defined by the vectors $\vec{n}_i$). For this sake we define the function $M( \vec{r})$ such that
\begin{equation}
\label{m_de_r}  M( \vec{r}) := \frac{1}{N} \sum^N_{i=1}|\vec{d_i}(\vec{r},\vec{n}_i)|^2, 
\end{equation}
where $|\vec{d}|$ is the distance of point $\vec{r}$ to the line defined by $\vec{n}_i$ (see Fig. \ref{d_definition}). The center of curvature is defined as the point $\vec{r}_0$ that minimizes the function $M( \vec{r})$.

If the object has a well-defined curvature center, the function $ M( \vec{r})$ not only has a clear global minimum, but small deviations from $r_0$ also lead to a large increase in $ M( \vec{r})$.
We may treat $ M( \vec{r})$ analogously to a chi-squared function and define a confidence region (CR) such that
\begin{equation}
\label{RC}  M( \vec{r})-M_0 \leq \sigma_m , 
\end{equation} where $M_0 = M( \vec{r}_0)$ and $\sigma_m$ is an arbitrary parameter. In this paper we use $\sigma_m=1$. 

We expect that curved arcs have a small CR as compared to the arc size. We noticed by visual inspection of the CR and the objects, that for arcs, the CR is usually elongated along the radial direction and does not in general intersect the arc.

We provide, as inputs to the ANN, combinations of the parameters described above derived from the Mediatrix filamentation. These combinations are defined so as to be scale invariant, such that they depend mostly on the object shape and are weakly sensitive to the pixel scale. In some cases we normalize the output by the appropriate power of $L$ to produce the scale invariant quantities.
In particular we tested the ANN with the following set of parameters:
 
$i$) The length-to-width ratio $L/W$.

$ii$) The mean of the scalar products of each unitary vector $\vec{n}_i/|\vec{n}_i|$ with its neighbor, $\vec{n}_{i+1}/|\vec{n_{i+1}}|$, i.e.,
\begin{equation}
\label{s}  s:= \frac{1}{N} \sum^{N-1}_{i=1} \frac{\vec{n}_i}{|\vec{n}_i|} \centerdot \frac{\vec{n}_{i+1}}{|\vec{n}_{i+1}|}. 
\end{equation} 
This quantity provides a measurement of the coherence of the shape. For very irregular objects, its value is low, while for long and smooth objects (curved or not), its value should be close to 1. 
  
$iii$) The minimum value of the function $M( \vec{r})$ divided by the arc length squared (for dimensional reasons), $M_0/L^2$. 

$iv$)  
The arc aperture $\Delta \theta:=L/R$, where the radius $R$ is the distance from $\vec{r}_0$ to $M^1_1$.

$v$) The ratio between the arc aperture defined above and the one constructed from the circle that contains points $E_1$, $M^1_1$, and $E_2$,
$\Delta \theta/\Delta \theta_c := \left(L/R\right)/\left(L_c/R_c\right)$. 

$vi$) The distance between the center of the circle $\vec{r}_c$ and the minimum of $M(\vec{r})$ normalized by the arc length,
$\delta r: = |\vec{r}_0-\vec{r}_c|/L.$

$vii$) The ratio between the major axis of the CR, $L_{CR}$, and the arc length: $L_{CR}/L$.

$viii$) The eccentricity of the CR, $e_{\rm CR}$.

$ix$) The ratio between the number of pixels from the arc enclosed by the CR, $A_{CR}$, and total number of pixels in the object, $A$, i.e., 
$A_{CR}/A$.

The choice of parameters above is somewhat arbitrary, but is inspired by the visual assessment of these quantities on samples of arcs and objects that are clearly nonarcs. An important component of this paper is to 
obtain a set that is at the same time good for discriminating arcs from nonarcs and is less time consuming.

\subsection{\label{RNA} Arc identification with an artificial neural network}

The arc identification process through the ANN can be subdivided into two parts: the training process and the actual classification. In the current implementation
we use a standard back-propagation and fully connected ANN~\citep{williams1986learning,rumelhart1988learning}. The ANN has the following structure: {\it a)} an input layer with $i$ neurons, where $i$ is the number of inputs used in the specific ANN configuration, which in this case is
a subset of the parameters described in section \ref{mediatrix}; {\it b)} a second layer with $j$ hidden neurons;
and  {\it c)}  the output layer with one neuron. 
The ANN activation function is linear and the output is a floating point number $R$ in the range  $-1$ to $1$.
The AMA code was developed using the python binding for the Fast Artificial Neural Network (FANN) library\footnote{For further information see \texttt{http://leenissen.dk/fann/wp/}}.   

In order to recognize the arc shape using this type of ANN, it is necessary to train this neural network on a group of objects, which were previously classified as arcs and nonarcs.
A successful training process is determinant to reach acceptable results in any back-propagation ANN code. The training requires presenting to the ANN two groups: the arcs group (AG), with desired output $+1$ and the nonarc group (NAG), with desired output $-1$. The two groups also need to have the same order of number of objects, otherwise the ANN may just output as the result a number that represents the larger group. 
The AG and NAG are split into a training group and a validation group. In this work we used $80\%$ from all objects chosen randomly from the sample of AG and NAG for training and $20\%$ for validation.

After the training, the ANN is applied to the validation group, yielding an output value $R$ for each object. 
We have thus to set 
a threshold $t$ on this output such that the code finally 
classifies each object (i.e., each set of input measurements on the object) as an arc or not.
After running in the validation group  
the code computes the completeness, $c$, defined as the fraction of arcs recovered (i.e., {\it the ratio of the number of detected arcs and the total number of arcs} in the validation group) and the fraction of false positives, $f$, defined as the fraction of nonarcs that are classified as arcs (i.e.,  the ratio of the number of nonarcs detected as arcs and the total number of nonarcs in the validation group). 

After the validation test, 
the training and validation groups are redefined randomly and the whole process is repeated $k$ times. In this work we retrained the ANN for a single set of input parameters $40$ times. The validation code outputs the mean completeness $\bar{c}$, mean false positive fraction $\bar{f}$, and their standard deviation for the $k=40$ validation groups.

\section{\label{train_and_validation} Training and validation of the ANN}

In this work we use the training and validation steps to characterize the behavior of the AMA with respect to all aspects of the ANN identification process mentioned in Sec. \ref{RNA} above, including the types of images used in the training, sets of inputs, number of hidden neurons, and threshold for classification. 
The goal is to define a good set of configurations for practical applications of the AMA.
In particular we seek to have a high completeness $c$ at the same time limiting the fraction of false positives $f$.
This search for the best parameters for the ANN detection is described in Sec. \ref{training}

The arcfinder method was trained using a sample of $175$ simulated arcs (AG) described in Sec. \ref{addarcs} and $437$ nonarcs (NAG) taken from HST images, as described in Sec. \ref{aplication}.
These numbers are the result of steps 1 (object identification) and 2 (preselection), and thus all objects from the two samples already pass the preselection criteria described in Sec. \ref{identificacao}. In particular all have ellipticities above  $e_{\rm th}=0.4$.
Therefore, in all comparisons and tests described in this paper we are really testing the measurements + ANN steps of the whole arc finding process, which is the aim of this contribution.

\subsection{\label{paramsinputs} ANN inputs from the Mediatrix filamentation}

 We performed the training with ten different subsets of the parameters {\it i} to {\it ix} described in Sec. \ref{mediatrix} to determine the best combination of Mediatrix parameters to be used as input
for the ANN.
Each subset is labeled with a letter from A to J. 
The input configurations are presented in Table \ref{combinations}.
We divided the input parameter sets into two groups depending on whether the CR evaluation is necessary or not for a given configuration. Group $1$ includes only measurements derived directly from the Mediatrix filamentation process ({\it i} and {\it ii}) and from the minimization of $M(\vec r)$ ({\it iii} to {\it vi}), while group $2$
contains measurements that depend on the CR (parameters {\it vii} to {\it ix}).
The total time to run the AMA varies only slightly within each group, but changes considerably between the two groups, as the process to obtain the CR is currently the most time consuming step of the AMA.

From sets A to J, the number of inputs (i.e., the number of parameters in the input vectors) is systematically decreased (except for I and J, which have only one input each).  
The three sets of configurations in Group 2 (A, B, C) include the determination of the CR and are thus those that take more computational time. From configuration D downward the time drops substantially. In all cases but one (J), we keep the parameter ($i$), i.e., $L/W$, which is historically the primer arc indicator.
In Sec. \ref{training} we test the AMA for each set and compare the results for $c$ and $f$ to define the best set for practical applications, both in terms of maximizing completeness and minimizing contamination, also accounting for the computational time.

\begin {table*}[htp]%
\caption{Combinations of inputs used for the neural network training (A to J).} 
\label{combinations}
\begin {tabular}{c|ccc|cccccc}
\toprule
& Group & 2 &  & Group & 1 \\
\hline
& $L_{CR}/L$  & $e_{\rm CR}$  & $A_{CR}/A$ & $L/W$   & $s$ &  $M_0/L^2$ & $\Delta \theta$ &  $\Delta \theta/\Delta \theta_c$ & $\delta r$  \\ 
\hline
A &X &X &X &X &X &X &X &X &X  \\

B & &X &X &X &X & X &X &X &X  \\

C &  &  &X &X &X &X  & X &X &X  \\

D &  &  &  &X &X & X & X & X &X  \\

E &  &  & & X  & & X & X & X & X  \\

F & & &  & X &  & & X& X& X  \\

G & & &  & X &  & &  & X& X  \\

H & & &  & X &  & & & & X  \\

I & & & & X &  & & & &  \\

J & & & & & & & & & X \\
\bottomrule
\end {tabular}

\end {table*}

\subsection{\label{addarcs} Arc simulations with AddArcs}

Given the intrinsic variation in gravitational arc shapes it is important to have a large enough training sample so as to encompass some of their diversity and, at the same time, have sufficient statistics to train the ANN. However, the current samples of arcs taken under uniform observing conditions are still substantially small. 
Also, we need a truth table of arcs in the AG and not all known arcs have spectroscopic confirmation. Moreover, we want to be able to control some observational and instrumental parameters, such as the background and noise, point spread function, and pixel size so as to test the arc finder under different conditions. For this sake, we use simulated gravitational arcs for the training and validation phases. 

We use exactly the same sample of simulated arcs as in \citet{BomMediatrix}.  That paper provides a more detailed description of the simulation process and the selection of this specific sample. Below  we provide an outline of the simulation procedure for completeness. 

The simulated sample was created using the \texttt{AddArcs} pipeline (Brandt et al., in preparation), which uses two input catalogs: one with the properties of the lenses (such as mass, ellipticity, and redshift) and one with the properties of the sources (such as magnitude, size, ellipticity, and redshift), plus a number of configurations that can be set, such as observational and instrumental parameters. The code distributes the sources in random positions for each lens in the catalog, following the specified surface number density, and then it randomly chooses the source parameters from the source catalog. Given the input models for the source and the lens, from their respective catalogs, the pipeline uses the \texttt{gravlens} code \citep{gravlens}
recursively to perform the projection of the sources onto the image plane. It then identifies which images correspond to arcs and generates postage stamps from them, providing as one of its outputs a pixelized surface brightness distribution of these objects, i.e., a simulated image of a gravitational arc. 

For our simulated arc sample the input catalog contains galaxy cluster scale halos from $N$-body simulations\footnote{We use a catalog from the  {\it Las Damas/Carmen} $N$-body simulation, \texttt{http://lss.phy.vanderbilt.edu/lasdamas/}.} and we assume a Navarro--Frenk--White density profile \citep[NFW;][]{nfw1,nfw2}, with elliptical surface mass density \citep[see, e.g.,][]{2013caminha}, a given mass--concentration parameter relation \citep{2008MNRAS.387..536G,2007MNRAS.381.1450N}, and fixed ellipticity. The sources are given by a S\'ersic surface brightness distribution \citep{sersic} with parameters derived from the Hubble Ultra Deep Field Survey \citep[UDF;][]{2006AJ....132.1729B,2006AJ....132..926C}. 

In Sec. \ref{aplication} we apply the trained AMA to real images of systems containing arcs taken with the WFPC2 instrument on HST \citep{2005Smith}. As in our training the AG is given by the simulated arc sample, we set the observational conditions in the simulations to mimic these HST images. In particular, we use the same pixel scale as WFPC2 and convert the counts in each pixel on the simulated images to data numbers using the properties of these HST images \citep[for details, see Appendix A of][]{BomMediatrix}. At this point the simulated arcs are smooth, i.e., the pixels have no fluctuations from noise and the simulated images have no background. We refer  to this calibrated set of arcs as pure arcs.

However, real astronomical images have noise (including Poisson noise from the counts in pixels) and background. Even though it is common to work on background-subtracted images, of course the background noise  remains. Therefore, for a proper test of the arc finding process we need to include at least these two effects on the simulated images, as they are of fundamental importance for object detection and measurement.
The (constant) background is added to all pixels as measured from the HST images. Then each pixel is assigned a new value sampled from a Poisson distribution with the mean given by that pixel value in count units (including object plus background). Finally, the new image with background and noise is converted again to the data units. For details of these processes, see Appendix A of \citet{BomMediatrix}. The derived images form the background and noise sample of arcs. 

Both samples of simulated arcs go through the object identification and preselection phases, as described in section \ref{apresentacao}. In particular, \texttt{SExtractor} is run on each simulated image containing one arc and a postage stamp is created for that object. The Mediatrix method is then applied and the derived parameters are input to the ANN.

The validation of the trained ANN is performed using the background and noise sample as it is the more realistic sample. Nevertheless, it is interesting to test the results of the training carried out 
using each type of image as entries. These tests are discussed in the next section.

\subsection{\label{training} The training and validation results: Determining good ANN configurations}

\begin{figure}[!ht]
 \begin{center}
 \sidecaption
 \resizebox{\hsize}{!}{\includegraphics{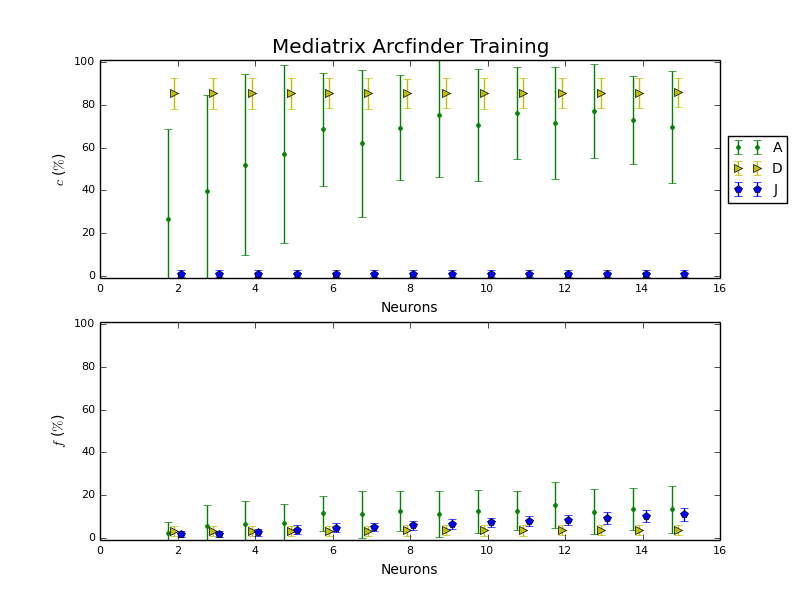}} 
 \caption{\label{neurongra} Mean completeness $\bar{c}$ and fraction of false positives $\bar{f}$ as function of hidden neurons, using background and noise arcs in the AG for training and a threshold $t=0$. The error bars are the standard deviation from the 40 training plus validation subsamples. We shifted the symbols horizontally for clarity.}
 \end{center}
 \end{figure}

In this section we present the training and validation results and use them to select an optimal set of configurations for the ANN arc finding process. We start by looking at the dependence of $f$ and $c$ as a function of the number of hidden neurons ($N_h$). We considered all configurations from A to J described in Table \ref{combinations} and varied the number of neurons from $2$ to $15$ fixing the threshold for the ANN output to $t=0$ (i.e., objects with the ANN response function $R>0 $ are defined as arcs). First we considered the training carried out on the pure arcs set and then on the background and noise images as the AG (the NAG is the same in both cases). In the first case, we do not notice any significant variation of the completeness and false positives with $N_h$ for any input configuration. On the other hand, when the training is performed using the images with background and noise, we do see a dependence on the hidden neurons for some of the configurations. This is shown in Fig. \ref{neurongra} for only three configurations:
A  (representative of group 2), D (representative of group 1 with several inputs), and J (with only one input). 
We see that the dependence of the completeness on $N_h$ is only significant for the A configuration, 
for which 
we can observe a significant increase in $c$ up to $N_h \sim 6$. 
In all configurations that we have investigated, there is no gain in increasing $N_h$ above this value. 
Thus, only for the configuration with the highest number of inputs and using the training set with more variation among the systems (due to the noise in the AG) does the ANN require more complexity than two hidden neurons.
On the other hand, the false positive fraction does increase with $N_h$ in most cases. This is less visible in configuration $A$ because of the large variance and  could be due to overtraining, when the number of neurons is large. 

As we see below, we end up choosing the pure arcs as a training sample 
and therefore we could be tempted to choose a very small number for $N_h$, as this also significantly decreases the computational time. On the other hand, to be on the safe side for real world applications, we still want to have a larger number of neurons  
for dealing with the diversity of real arcs.Therefore, we set our final number of hidden neurons to Nh = 4 in all tests to reach a balance between computational time, false positives, and completeness.
Therefore, we set our final number of hidden neurons to $N_h = 4$ in all tests to reach a balance between computational time, false positives, and completeness.

\begin{figure}[!ht]
 \begin{center}
 \sidecaption
 \resizebox{\hsize}{!}{\includegraphics{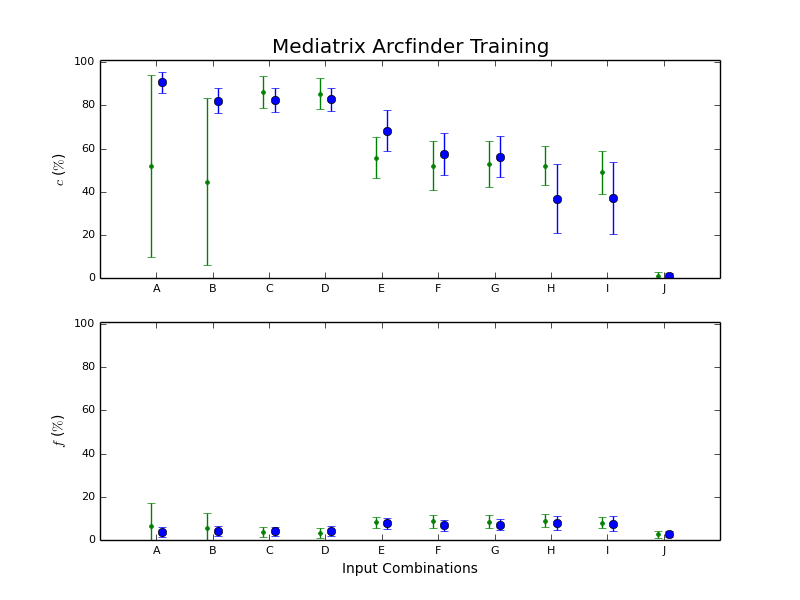}} 
 \caption{\label{configgra} Mean completeness $\bar{c}$ and false positive fraction $\bar{f}$ for the different sets of inputs (for threshold $t=0$). The results from the training in pure arcs  are shown as large blue dots, while those using images with background and noise are shown as small green dots. The error bars are the standard deviation from the sets of training plus validation subsamples. We shifted the symbols horizontally for the two types of input images for clarity.}
 \end{center}
 \end{figure}

Now we turn to the choice of the final set of configurations and the types of images for the training. In Fig. \ref{configgra} we show the results for $c$ and $f$ for the ten configurations in Table \ref{combinations} using pure arcs  (in large blue circles) and those with background and noise (in small green dots) for the training. Clearly, configurations A to D have the best performance, both for completeness and for false detections. The results from training on pure arcs in general have a smaller variance than using background and noise, in particular for configurations A and B. In addition, for lower thresholds, A and B have much more contamination when trained on arcs with background and noise than with pure arcs.

 At first sight it could seem surprising that training with the more realistic set of arcs in general gives worse results. For the ANN, it is better to learn with a more consistent and well-defined set of parameters from the arcs, than having a larger variation on these parameters owing to noise, even if the validation is carried out on the images that do have noise and background. It is important to mention that the noise is added only once to the arc sample, i.e., the variance among the 40 subsamples is not due to different realizations of the noise, but rather to the spread in the parameters caused by the noise in each subsample of arcs. For now on, we choose to carry out the training process using only the pure arc sample.

As for the inputs, we see from Fig. \ref{configgra} that the combination that gives the highest mean completeness and a low fraction of false detections is configuration A, from group 2. However, D also gives a good performance (as good as the other configurations in group 2) but is in group 1, i.e., is computationally faster, as it does not require the computation of the CR. We therefore keep these two sets of inputs, A (the best) and D (almost as good as A, for both $c$ and $f$, but faster), for the next test and for applications to real data.

\begin{figure}[!ht]
 \begin{center}
 \sidecaption
 \resizebox{\hsize}{!}{\includegraphics{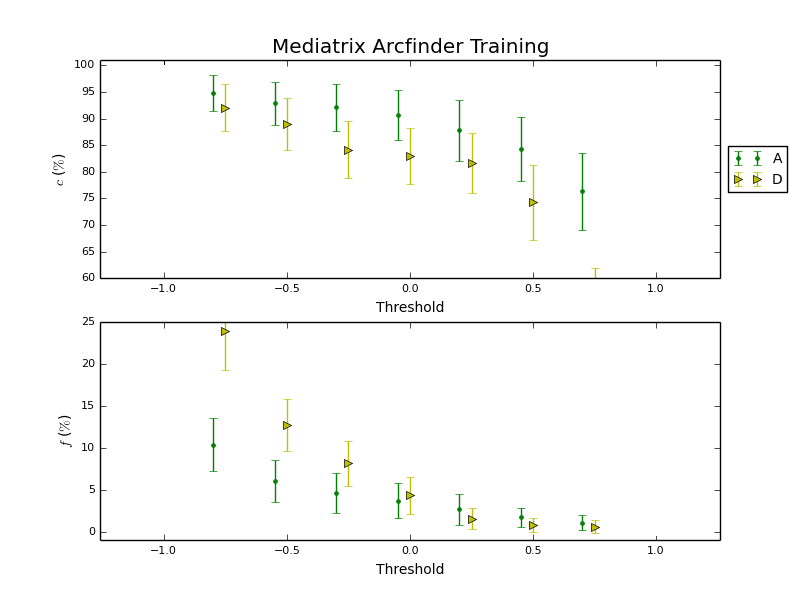}} 
 \caption{\label{thrgra}  Mean completeness $\bar{c}$ and mean false detection fraction $\bar{f}$ for the input configurations A and D as function of threshold. The results are obtained from the training on pure arcs with an ANN with four hidden neurons. Symbols are shifted in the horizontal direction for clarity.}
 \end{center}
 \end{figure}

Finally we look at the dependence of $c$ and $f$ on the threshold $t$ for these two selected configurations.
We vary $t$ in steps of $0.25$. The cases $t=-1$ and $t=+1$ are trivial as all objects are classified as arcs and nonarcs, respectively. In Fig. \ref{thrgra} we show the results  for $t$ in the range $[-0.75,0.75]$. 
As expected, both the completeness and the false detections decrease as $t$ is increased. However $c$ has a softer dependence with $t$ than $f$.
Two possible choices for $t$ are in order. If we want to have a higher completeness, even at the expense of a higher percentage of false detections, then  $t=-0.75$ is a good choice. This threshold would be preferred, for example, in targeted surveys, where a visual inspection to discard false positives is feasible even if the false detections outnumber the real arcs. 
In this case we obtain $c \sim 95 \%$ and $ 90 \% $ and $ f \sim 10\%$ and  $25\%$ for configurations A and D, respectively.
On the other hand, if we seek a purer sample of arcs, a good choice is $t=+0.25$. After this value $c$ drops considerably, while $f$ does not vary much. This choice could typically be adopted for a wide-field survey, where we need to minimize the fraction of the objects to be inspected for a final selection. In this case $c$ decreases a bit to $\sim 90\%$ and $80\%$, but $f$ drops substantially to $\sim 3 \%$ and $2 \%$, respectively, for A and D.

We recall that $f$ is defined as the fraction of false positives with respect to the total number of nonarcs, i.e., it is essentially the number of false detections over the total number of objects that pass the preselection cuts. In the training and validation process the numbers of arcs and nonarcs are of the same order of magnitude. However, for wide-field surveys, the number of arcs is roughly five orders of magnitude less than the total number of objects detected. Therefore, even if the preselection phase filters out $90\%$ of the objects and $f$ is as low as $1\%$, the false detections would still outnumber the real arcs by large amounts. Thus, even a low contamination as currently achieved with $t=+0.25$ would still require a further step of visual inspection when applied to large surveys, as happens with most arc finders proposed so far.

It is worth pointing out that $t$ can always be set a posteriori in the sense that the ANN is specified without the need to define a threshold. Once the inputs and hidden neurons are defined and the training is carried out on a given sample, the ANN is fully determined. When the ANN is applied to the data, the result is an output value of the response function for each object. Therefore we can vary the value of $t$ after the ANN is run and choose a suitable balance between $c$ and $f$ to set the threshold.

We see from Fig. \ref{thrgra} that the completenesses for configurations A and D are compatible with each other
within their standard deviations for the whole range of $t$ (except for $t =0.75$),
showing that both configurations are comparable for arc detection (although $\bar{c}$ is systematically higher for A).
Regarding the false detection fraction, it  is clearly higher for configuration D and $t<0$. The highest difference with A occurs for $t = -0.75$ and is smaller than two standard deviations. It is not clear whether this is a real difference between the two configurations or if it is just a fluctuation. 

In the next section we apply the two ANN (i.e., with configurations A and D) trained as described in this section to objects from real HST images.

\section{\label{aplication}  Application on HST cluster images}

In this section we show an example of application of the AMA to real images taken with the HST. In particular, we consider a well-known sample of massive clusters observed with 
the Wide Field and Planetary Camera 2 (WFPC2) instrument on HST from \citet{2005Smith}. 
Images from this camera have been used in other exploratory studies of arc finders \citep{2004Lenzen,2005Horesh,2007Seidel}, and, in particular, the same \citet{2005Smith} sample was used in \citet{2005Horesh}.

The WFPC2 instrument has a mosaic of three wide-field (WF) CCDs (forming an L pattern) and a smaller CCD with finer pixel scale close to the center of the field.
The exposures of each cluster are centered on one of the WF CCDs. Several exposures are taken with a dither pattern, so as to combine all CCDs into a single image with no gaps. The HST server%
\footnote{The HST data products can be downloaded from the European HST Archive at ESA/ESAC: 
\texttt{http://archives.esac.esa.int/ehst/}}
provides both the combined image with all CCDs and a combined image of all exposures for each single CCD. 
For the purposes of this paper it is better to work on the single images per CCD, as the mosaic images have strong $S/N$ variations and artifacts 
in the regions between CCDs and close to the edges. This can produce spurious detections and affect the background estimation and it is beyond the scope of this paper to deal with them. For each CCD image we remove the regions near the image borders to avoid the spurious detections.

We use the images from \citet{2005Smith} 
to apply the AMA to find arcs in the images, but also to provide the sample of nonarcs for the training of the ANN. This is so that the NAG have exactly the same observational and instrumental conditions as the images in which we look for arcs. We mimic those same conditions in one of the simulated arc samples as discussed in\footnote{ For details on the noise and background evaluation in the HST images 
see \citet{BomMediatrix}.}
Sec.\ref{addarcs}. In particular, we consider CCDs that do not contain the cluster center and have no apparent arcs to provide the NAG. We use seven such CCDs, 
carry out the detection and preselection steps and end up with the sample of 437 nonarcs used in the training of the ANN discussed in the previous section.

\begin{figure}[!ht]
 \begin{center}
 \sidecaption
 \resizebox{\hsize}{!}{\includegraphics{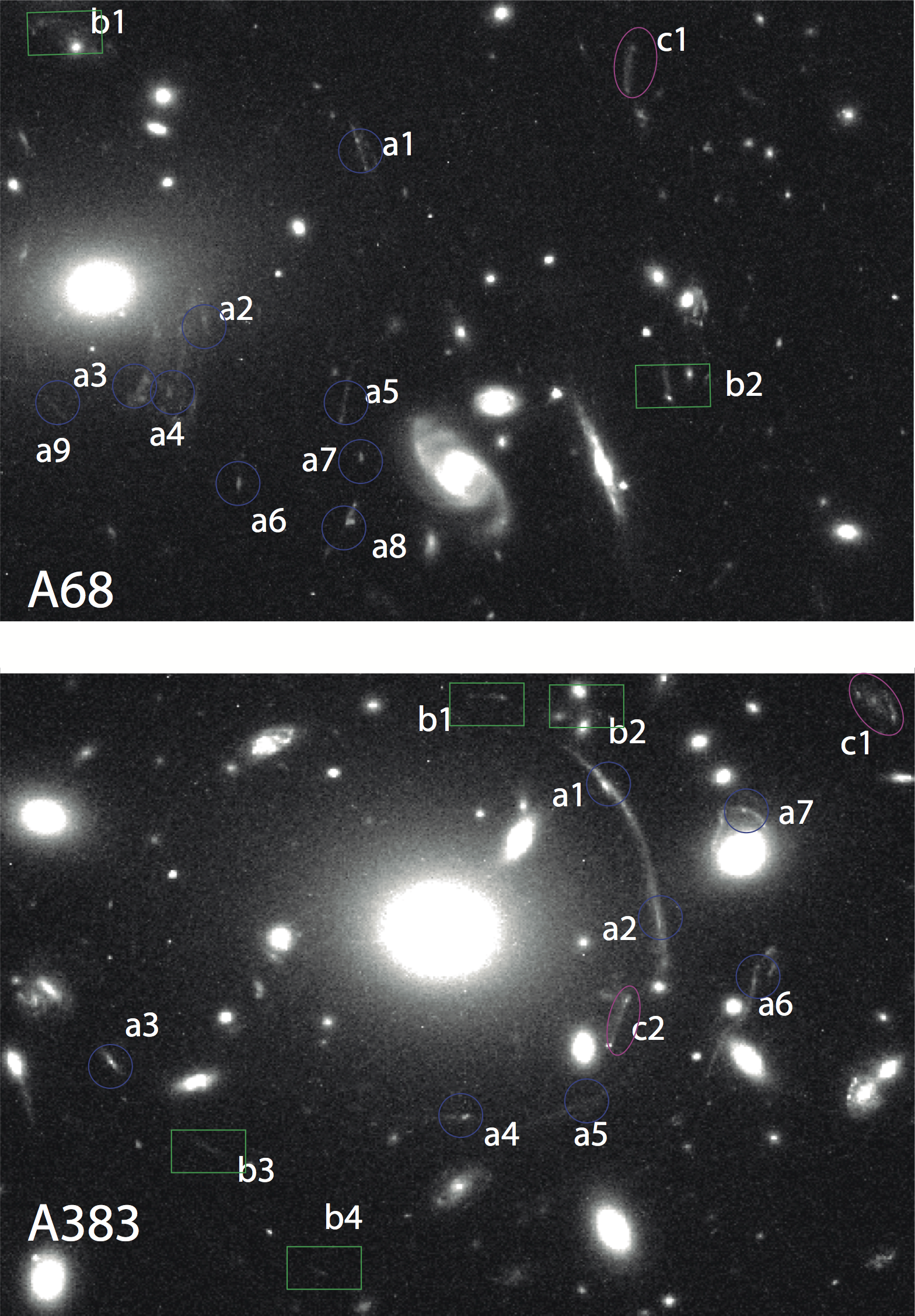}}
 \caption{Arc candidates in Abell 68 and Abell 383. The candidates were classified in $3$ categories: A, best candidates indicated with blue circles; B, intermediate candidates indicated with green rectangles; and C, ambiguous candidates indicated with magenta ellipses.}
 \label{A68_and_A383}
 \end{center}
 \end{figure}

\begin{figure}[!ht]
 \begin{center}
 \sidecaption
 \resizebox{\hsize}{!}{\includegraphics{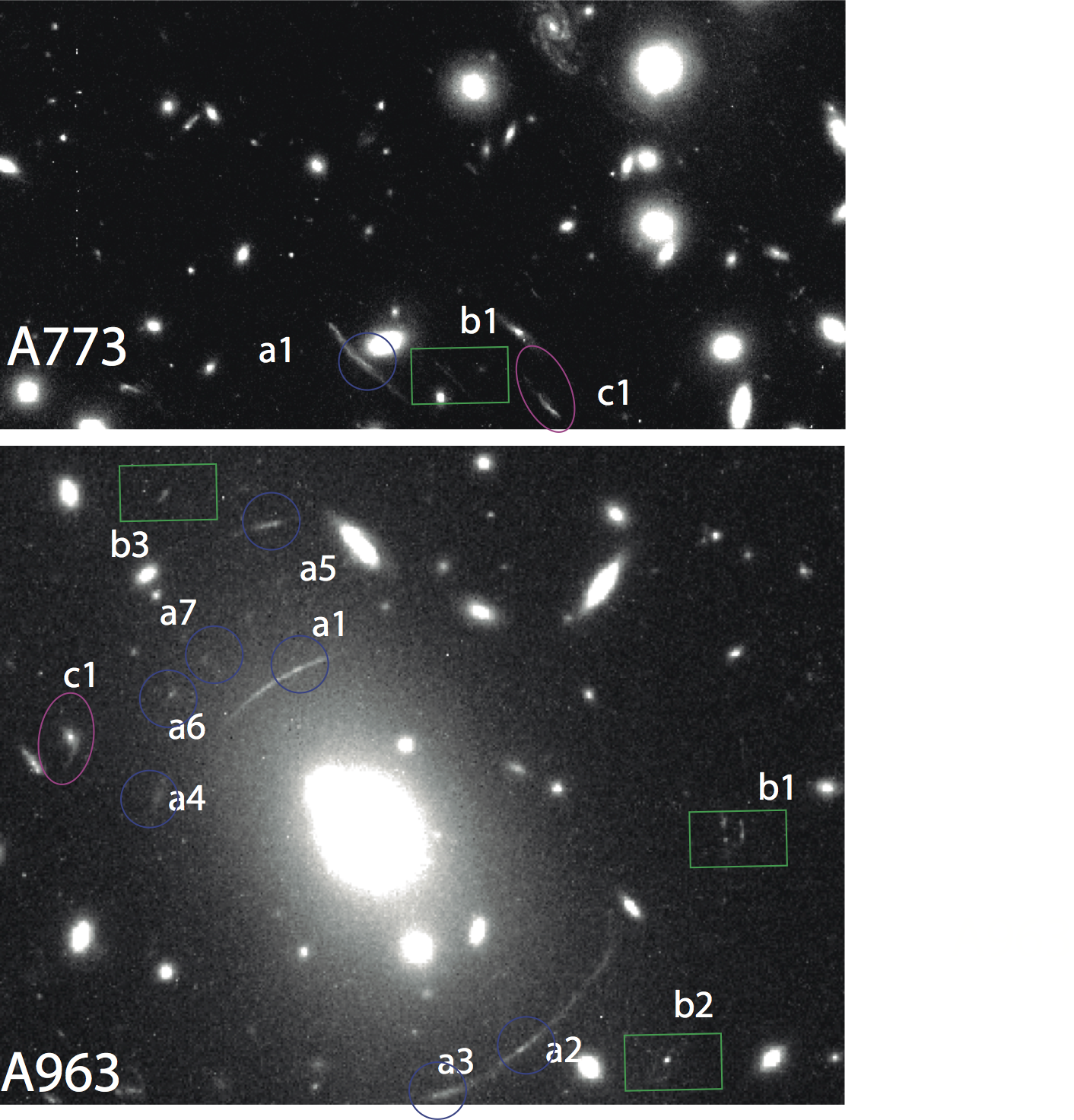}}
 \caption{Arc candidates in Abell773 and Abell963. The arc candidates in the $3$ categories are indicated following the same convention as in Fig. \ref{A68_and_A383}.}
  \label{A773_and_A963}
 \end{center}
 \end{figure}

We apply the AMA to four clusters in the  \citet{2005Smith} sample that have giant and clearly visible arcs, namely Abell 68, Abell
383, Abell 773, and Abell 963.
The images of the CCD with more arcs for each cluster were visually inspected and the arc candidates were classified in three categories:
A for the best candidates, objects with curved shapes close to the cluster central regions or galaxy cluster members; B for intermediate candidates, which are curved but do not have a cluster center or  galaxy as a center of curvature or that are close to the cluster center but are not curved; and C for more ambiguous candidates that do not fall in the previous categories. This classification is somewhat arbitrary
but is useful for a first assessment of the ability of the ANN to recover the arcs as a function of their quality/likelihood. For each cluster we label the arcs in each category with a number (e.g. a4, b2, and c1). In Figs. \ref{A68_and_A383} and \ref{A773_and_A963} we show cutouts of the images encompassing the regions of each selected CCD where arcs were visually selected and marked.
In cluster A68 we marked 9, 2, and 1 arcs in the A, B, and C categories, respectively. For A383 we labeled 7, 4, and 2 arcs in these categories, while for A773 the numbers are 1, 1, 1, and for A963 7, 3, 1. This gives a total of 39 arc candidates, 24, 10, and 5 respectively in the A, B, and C, categories. We use these identifications as a truth table for testing the AMA.

The aim here is only to have a set of objects with a morphology visually associated with gravitational arcs.
It is hard to compare the numbers above with the other identifications for the same clusters in the literature. On one hand, we only consider the arcs from a single CCD, do not include radial arcs, and do not impose an $L/W$ cut for arc selection. On the other hand the visual identification is rather ambiguous anyway.
In any case the orders of magnitude are compatible with the visual searches in  
\citet{2005ApJ...627...32S}, which identified $27$ arc systems in these four clusters and \citet{2005Smith}, which found $33$ multiple images in the four CCD chips we consider in this work.

Now we apply the AMA algorithm, following the steps described in Sec. \ref{apresentacao}, to the four selected clusters. First we run 
\texttt{SExtractor} (step 1), with the same configurations used in the training and validation phases, obtaining a total of 1378 detections. Applying the ellipticity and other cuts on the preselection (step 2) leaves us with 304 objects, on which we run the Mediatrix filamentation (step 3), providing the inputs for the ANN.
It must be pointed out that not all arcs visually identified comprising our truth table are found to be objects in the \texttt{SExtractor} run (likely because of their low surface brightness). Moreover, many of the arcs found end up blended with other objects in the image and therefore the morphology of the generated postage stamp does not represent an arc anymore. The total number of arcs that are either not detected or blended is 15. 
For a fair comparison with the results of Sec. \ref{training} these objects should not be considered in the denominator for computing the completeness $c$, as we do not expect the ANN to identify them as objects; the ANN was not trained on blended objects and obviously the arcs that are not detected cannot be classified by the ANN. 

Finally, we apply the ANN, trained as described in Sec. \ref{training}, using either the sets of inputs A or D, which are selected from our exploratory runs on the simulated arcs (using four hidden neurons and trained using the pure arcs as AG). The resulting number of arcs detected is of course a function of the threshold $t$. In Table \ref{resultsreal}, as an
example we show the results using configuration D and $t=-0.75$. In that case, a total of 16 arcs were identified and there were 43 false detections.  
We also show the number of arcs visually identified in each category along with the arcs that were not detected by \texttt{SExtractor} and those that were blended with other objects. We note that in A68 four out of six detectable arcs were found, while for A383 six out of seven were identified by the ANN, in A773 one out of two and in A963 five out of eight. The overall $c$ in this case is $16/24\simeq 67 \%$. 
Considering each subclass, we have a completeness of $71\%$, $67\%$, and $50\%$ 
for the A, B and C categories respectively. Although these numbers show the expected trend with arc quality, they are all mutually consistent taking Poisson statistics into account, and we cannot conclude whether the arc finder performs better or not with arc quality. Therefore, in the remaining of this paper we consider all categories together for evaluating $c$.

\begin{center}
\begin {table*}[htp]%
\caption{Arc detections in the 4 selected HST clusters. Columns 2 to 6: number of arc candidates visually identified in each category (A, B, and C), objects that are not detected by \texttt{SExtractor,} and arcs that are blended with other objects. Column 7 shows the arcs detected by the ANN for configuration D and $t=-0.75$. The last column shows the number of false positives (i.e., objects classified as arcs but not on the arc truth table) for the same configurations.}
\label{resultsreal}
\begin {tabular}{|c |c | c | c | c | c | | c |c |}
\hline \hline 
 &  \multicolumn{3}{c|}{Category} & \multicolumn{2}{c||}{SExtractor} & \multicolumn{2}{c|}{ANN}\\
\hline \hline  
Cluster & A &B & C &  Not detected & Blended  & Arc detections & $N_f$   \\
\hline 
Abell 68 & 9 & 2 & 1 & a9 & a2, a3, a4, b1 & a6, a7, a8, b2 &$12$ \\  
\hline
Abell 383 & 7 & 4 & 2 &  a5, b1, b4 & a7, b2, c2 & a1, a2, a5, a6, b4, c1 & $3$ \\
\hline
Abell 773 & 1 & 1 & 1& --- & a1 &  c1 & $13$ \\
\hline
Abell 963 & 7 & 3 & 1 & --- & a1, a3, a6 &  a2, a5, a7, b2, b3 & $15$ \\
\hline \hline 
\end {tabular}
\end {table*}
\end{center}

As a comparison, we point out that the arc finder run presented in \citet{2005Horesh} found 16 arcs in the four clusters under consideration. Restricting to the areas where we denoted arcs for our truth table (see Figs. \ref{A68_and_A383} and \ref{A773_and_A963}) these authors found a total of $9$ arcs, $5$ of which are in common with the sample of 16 arcs detected with the AMA, and $3$ are blended in our detections.
On the other hand, only objects with $L/W > 7$ are selected by  \citet{2005Horesh}, while we make no cuts in this quantity. In any case, the focus in  \citet{2005Horesh} is not on completeness, but on a comparison between a real and a simulated arc sample. 
A more detailed comparison of the AMA with this and other arc finders is outside the scope of this paper 
\citep[see][for preliminary results]{2015mgm..conf.2088D}. 

As for the false detection fraction $f$, it is computed as the ratio of the false positives
to the total number of nonarcs given as input to the ANN. 
For the configuration in Table \ref{resultsreal} we have $f \sim 15\%$. With respect to the total number of objects detected, the fraction of false detections is $3\%$.

The false positives are objects classified as arcs, but that are not on the truth table. In principle, some of these objects could be real arcs that are missed by visual inspection. However, we did look at all false positives and only two of them could be associated with arcs; they were in fact pieces of arcs with multiple peaks, other pieces of which have been identified by the arc finder. Thus these cases are negligible for the purposes of this paper.

\begin{figure}[!ht]
 \begin{center}
 \sidecaption
 \resizebox{\hsize}{!}{\includegraphics{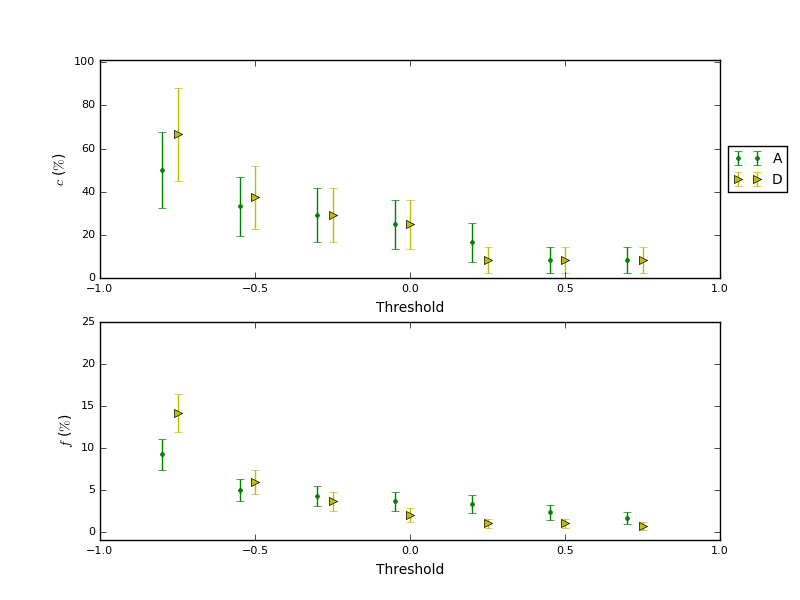}} 
 \caption{\label{cfHST}  Completeness $c$ and fraction of false detections $f$, including Poisson error bars, for the arcs in the 4 selected HST Abell clusters, for configurations A and D, as a function of threshold.
Data points for configuration A are shifted horizontally for clarity.}
 \end{center}
 \end{figure}

In Fig. \ref{cfHST}  we show $c$ and $f$ as a function of the threshold $t$ for configurations A and D combining the arc detections  for the four cluster images considered in this section. Poisson error bars are indicated. We see that the ANN achieves a reasonable completeness, $\sim  70\%$ for configuration D, and  $\sim  50\%$ for A, for a low threshold ($t= -0.75$), while still keeping a low contamination rate, $\sim 15\%$ and $\sim 9\% $ for configurations D and A, respectively. 

As in the simulations, configurations A and D yield values of $c$ that are compatible with each other in the whole range of $t$, taking their Poisson errors into account.
However,  $c$ differs considerably between the real and simulated data. The completeness is significantly lower on the data and decreases more abruptly with the threshold than what was seen in the simulations.
We interpret this result as an indication that the simulations are not realistic enough for a proper quantitative comparison with the real data. In fact, 
the simulated arcs are very diverse, but do not include some relevant degrees of realism, such as surface brightness variations in the sources and lenses with substructure. In addition, the real arcs are often close to bright galaxy cluster members, which can affect their segmentation and deblending (both due to contamination from the galaxies and background misestimation on the crowded field) and thus affect their shape. The strong effect of blending with other objects is already accounted for in our comparisons in the sense that these objects are removed from the denominator of $c$. However, a less significant contamination from close-by objects or a breaking into smaller objects affects the Mediatrix measurements and therefore the arc detection with the ANN, which was trained in simulations that do not include these effects. 

In the validation $c$ is a bit higher for A than D in the whole range of $t$.
On the other hand, in the real data there is an apparent trend for D to have a higher completeness than configuration A for the smallest threshold. For the remaining interval the performance is very similar among configurations A and D.

Regarding $f$, the results are also compatible between A and D within the error bars, and the difference is again higher for $t=-0.75$. The false positive fraction drops substantially for higher $t$, reaching $f \lesssim 5\%$ for  $t>-0.25$.
 
Comparing the false positive fraction obtained during the validation process with those from the runs on the HST images (bottom pannels of Figs.  \ref{thrgra} and \ref{cfHST}), we see that the results are consistent for configuration A in the whole range of $t$. 
However, for configuration D and $t<0$ the false detection fraction is clearly higher in the validation than in the real data. The highest discrepancy occurs for $t = -0.75$, but is smaller than twice the standard deviation within the training plus validation samples.
This difference could be just a consequence of the difference in $f$ between A and D pointed out in the previous section or may be a difference in the behavior of $f$ in the real data with respect to the validation.

The agreement between the validation set and the real data for $f$ is expected; 
the NAG is obtained from the same set of images from WFPC2 for the training plus validation, although the CCDs are different and include other clusters. Thus we would expect a similar behavior between the bottom panels of Figs. \ref{thrgra} and \ref{cfHST} for $f$. This is indeed the case for configuration A and strengthens the case for a fluctuation in the false positives obtained in the simulations for configuration D.

\section{Discussion}  \label{discusao} 

The purpose of the paper is to present the AMA%
\footnote{The method is implemented in the python language and the source code is available upon request to the authors. 
The training sample is also available upon request.}
and provide a simple example of application, first illustrating the training and validation process on simulated arcs and then the application to a real, albeit small, data set.
The major novel aspects of the present work are the use of the Mediatrix method in step $3$ of the arc finding process and the use of simulations. The simulated arcs are used not only to train, but also to find a good set of ANN configurations for step $4$. 
There is room for many improvements in the processes described in this paper, most notably in the use of more realistic simulations and increasing the number of systems in the application to real data, but also in other aspects arc detection. 

On the object identification 
and segmentation side, betterments can be implemented in order to detect faint sources, avoid the breaking of large arcs, improve on the deblending with nearby objects, and to find and segment sources in high background regions, which are known issues for arc identification.
Although \texttt{SExtractor} does not deal with all these issues in an optimal way for arc detection, it has been used in many arc finders for their object identification to some degree \citep[e.g.,][]{2005Horesh,2007Estrada,Kubo2008,Marshall2009robot,JosephPCA2014, Maturi2014}.
A better performance than running \texttt{SExtractor} in a straight way, as in the current work,
has been obtained by carrying out multiple runs of this code with different thresholds \citep{2005Horesh}, or by using \texttt{SExtractor} alone for pixel thresholding \citep{Kubo2008}.
Other arc finders use different approaches for object detection and segmentation, which are specifically oriented toward identifying arcs \citep[e.g.,][]{2004Lenzen,2006Alard,2007Seidel,2012More,2016ApJ...817...85X}.

A key issue for detection and segmentation for arcs is that these objects are often embedded in the haloes of bright galaxies (especially for galaxy-scale arcs and radial arcs) or blended with foreground galaxies (especially for arcs in clusters). One approach that has been implemented to address this issue is to fit and subtract the light profile of galaxies, as in \citet{2005ApJ...627...32S,Brault2015}. Several 
codes have been proposed in the literature to this end \citep[e.g.,][]{PyMorph2010,GALPHAT2011,GALAPAGOS2012}, often running \texttt{galfit}~\citep{galfit32010} recursively to fit each galaxy by a combination of elliptical brightness distributions with \citet{sersic} profiles. Advanced versions of \texttt{SExtractor} also fit and subtract all identified objects in a field \citep[e.g.,][Moraes et al., in prep.]{SEMF2012ApJ...757...83D,SEMF2015A&A...578A..79D}. Other schemes to subtract objects from images, which could be useful for arc finding have also been proposed \citep[e.g.,][]{JimenezTeja2012}.
A different approach has been carried out by \citet{2016ApJ...817...85X}, who propose a new detection and segmentation scheme, working in intensity difference space, which performs well in bright halos without the need of subtraction.

Regarding the preselection, for applications to wide-field surveys this phase must also include the removal of image artifacts such as  satellite tracks, star spikes, and regions with a large amount of noise or with a steeply varying background. In the current example we remove noisy regions by simply cutting off objects that are close to the CCD borders. However the AMA already includes a proper handling of survey masks, which are produced to avoid bright star halos and spikes, satellites, and other image features. The approach of \citet{2016ApJ...817...85X} is also well suited to remove diffraction spikes without the need to use masks and may be useful for less bright nonmasked stars whose spikes could contaminate the arc detection. 

For object measurement we propose the use of the Mediatrix filamentation method, and several parameters derived from it, as it was designed for elongated and curved objects. Most arc finders end up using less parameters and simpler measurement schemes to characterize the arc candidates, such as $L$ and $L/W$,
and only in a few cases include estimates of the curvature \citep{2007Estrada,Kubo2008}. However, other sets of inputs in addition to the Mediatrix inputs could be given to the ANN, such as higher order moments of the brightness distribution, including the arcness \citep{Kubo2008}. 

We argue that by using a machine learning algorithm for the final candidate selection (in this case a back-propagation ANN) one may achieve a better efficiency in finding arcs than using hard cuts on a few variables. By working on a multidimensional parameter space, it is possible to deal with correlations among the variables and empirically obtain combinations that represent gravitational arcs. For example, arcs tend to be more curved and smaller at galaxy scales  than in massive clusters, such that a single cut in $L/W$ or arcness would not be optimal for finding arcs at both scales.

Artificial neural networks were first used by \citet{2007Estrada} to search for arcs. In their case the simulated arcs are simply sections of a circle with a 
surface brightness profile that is uniform along the tangential direction and is convolved with a Gaussian with FWHM similar to the typical seeing of the images. The ANN is trained using a hundred such simulated arcs covering a range of sizes and brightnesses, which are added to SDSS images.
The objects are also identified with \texttt{SExtractor} and the preselection is also carried out using an estimate of the object's elongation. Finally, four inputs are given to the ANN, based on a fit of the object by a circle and on a determination of the object's length (using its furthest pixels).
\citet{2007Estrada} study the efficiency for recovering the simulated arcs both for a visual inspection and for the automated process as a function of  peak surface brightness and $L/W$.
A maximum efficiency of $40\%$ (with respect to the number of simulated arcs) is achieved  in the automated search. The present work can be seen as an improvement on \citet{2007Estrada} in the sense that we use more realistic simulated arcs and a wider set of input measurements 
well suited to characterize the arcs, in addition to tuning the ANN configurations to improve the completeness. 

Of course the key to a good performance in a learning algorithm is the realism of the training sample. Many improvements can be incorporated into the simple \texttt{AddArcs} simulations described in this work, such as considering more realistic lenses \citep[e.g.,][]{2011MNRAS.418...54H,2016ApJ...817...85X} and sources \citep{Kubo2008,Marshall2009robot, 2011MNRAS.418...54H}. 
Moreover, in addition to having a realistic sample of isolated arcs, those have to be added to real images, for example,  to address the issue of blending with other sources and of embedding in the halo of bright galaxies in cluster cores.
Other works have used simulations to test arc finders, define their parameters, or train their methods, and in some cases determine the selection functions \citep[e.g.,][]{2005Horesh,Kubo2008,Marshall2009robot, 2011MNRAS.418...54H, RINGFINDER,JosephPCA2014,Brault2015,2016ApJ...817...85X}.
Another possibility is to use the growing number of strong lensing systems detected in wide-field surveys and HST images to perform the training on real data sets. For example, over 600 candidate systems have been detected in recent studies using CFHTLS data
\citep{Maturi2014,RINGFINDER,Brault2015,SPACEWARPSII,ParaficzCFHTLS}, which could be used to train and better characterize the AMA and other arc finders.
By training and validating the ANN with more realistic simulated arcs or with real data, we expect to 
reach a better agreement for $c$ in comparison to applications to other data sets (and therefore to achieve a higher completeness), which is different from what we found when applying our trained ANN to the HST data.

In general, the several arc finders proposed in the literature carry out an end-to-end approach from the science image to a list of arc candidates, implementing all four steps that we refer to in this paper. However, they have their own solutions for each step with different degrees of sophistication and specificity for finding arcs. For example, in this work we focus on the third and fourth steps, respectively, by using a set of object measurement parameters that are well suited to arcs and a trained ANN, while most methods use simple cuts on a few parameters for the final classification. On the other hand, we use a generic object segmentation code that is not optimal for arcs.
For most methods, these four steps could be performed interchangeably. Therefore, if they are presented in a modular way, we would be able to test the performance of each step independently.
Moreover, new arc finders could be created 
combining the solutions for each step from those available that work better in specific situations. 
Several possible concatenations of the arc finder modules could also be compared using their end-to-end performance. 
Given that about a dozen finders have already been proposed, and that we are on the verge of applying them on large data sets, it would be very useful to carry out such a comparison.
This is one of the avenues we will pursue for future work
\citep[see,][for preliminary results with a few arc finders]{2015mgm..conf.2088D}.

After a decade of progress in the development of gravitational arc finders, several codes
are ready for exploring the new generation of wide-field surveys in the quest for gravitational arcs.
However more progress is still needed for fully automated runs so as to produce samples that can be readily exploited for their applications.
Besides improving the efficiency in some situations, the most important is to limit the false positives to a level  low enough to be corrected for and certainly less than the number of real systems, as  thousands to hundreds of thousands strong lenses are expected 
in the forthcoming data.
Different strategies have been proposed and implemented to address these issues.
Combining aspects of these solutions, which are more suited to each step of arc detection,  seems a natural  way to proceed. We believe that using neural networks or other machine learning methods may provide an important contribution to the task of selecting more complete and pure samples of gravitational arcs for a broad range of deflector scales and backgrounds.

\begin{acknowledgements}
 C.~R.~Bom is partially funded by the Brazilian agency CNPq of the Ministry of Science and Technology and Innovation (MCTI).  
 M.~Makler is partially supported by CNPq (grant 312353/2015-4) and FAPERJ (grant E-26/110.516/2-2012).
C.~H.~Brandt is supported by the Brazilian agency CAPES, grant 15113/13-2 under the CAPES/ICRAnet agreement.
We thank J. Estrada, H. Lin, C. Furlanetto, B. Rossetto, A. More, B. Santiago,
 and the DES Strong Lensing Working Group members for useful comments along the development of this work.
This paper makes use of observations made with the NASA/ESA Hubble Space Telescope, obtained from the ESO/ST-ECF Science Archive Facility.

\end{acknowledgements}

\bibliographystyle{aa}
\bibliography{bibliografiaArcfinder}

\end{document}